\documentstyle[12pt]{article}
\begin{document}
\date{}
\title{The inverse scattering method for cylindrical
  gravitational waves}
\author{G. G. Varzugin${}^{\dag}$}
\maketitle
\begin{center}{\it $\dag$
Laboratory of Complex System Theory, Physics Institute of
St-Petersburg University, St. Petersburg, Peterhof, Ulyanovskaya
1, 198904.}\end{center} \vskip0.5cm

\begin{abstract}
The initial-value problem for cylindrical gravitational waves is
studied through the development of the inverse scattering method
scheme. The inverse scattering transform in this case can be
viewed as a transformation of the Cauchy data to the data on the
symmetry axis. The Riemann-Hilbert problem, which serves as
inverse transformation, is formulated in two different ways. We
consider Einstein-Rosen waves to illustrate the method.
\end{abstract}

\begin{section}{Introduction}

In this paper we study the cylindrical gravitational waves model
with two polarization modes. The space-time of the model possesses
two commuting one-parameter isometry groups of which one is
isomorphic to $R$ and the other to $SO(2)$, and is non-stationary
solution of the vacuum Einstein equations. The generators of these
groups are called Killing vector fields. In the case of one
polarization (both Killing fields are hypersurface orthogonal)
this model is also known as Einstein-Rosen waves. The theory of
Einstein-Rosen waves is linear, however, still physically
interesting and even in recent time provides us with new important
results both in classical and quantum scope
Ref.\cite{asym_structure,probing} where one finds additional
discussion and reference list. The polarized model is non-linear
and hence far more difficult to study.

In two Killing reduction of vacuum or electro-vacuum general
relativity Einstein equations belong to the class of integrable
equations. By the other words they can be written as a
compatibility condition of an auxiliary linear system. Although in
this paper we use the linear system proposed in Ref.\cite{B-Z} it
is worth mentioning that there exits essentially different
approach Ref.\cite{H-E-HomRHP}.

The best tool to study integrable equations is the Inverse
Scattering Method (ISM). There are many works where the ISM is
applied for constructing exact solutions and there are a few
papers where the physically relevant models are considered. The
author knows only Ref. \cite{H-E-plane-waves} where the model of
colliding gravitational plane waves was investigated and Rev.
\cite{mine} where the solutions with disconnected horizon were
analyzed.

The main goal of the present paper is to develop a framework to
study the cylindrical gravitational waves model.
\end{section}

\begin{section}{Field equations and boundery conditions}

The metric of the space-time with two commuting space-like Killing
fields can always be chosen locally in the following form
\begin{equation}
ds^2=ds_{II}^2+Xd\phi^2+2Wd\phi dz+Vdz^2, \label{metric}
\end{equation}
where $$ds_{II}^2=\gamma_{ab}dx_adx_b$$ is the metric of an
orthogonal to both Killing vectors two-surface. The variables $ X,
W, V$ depend only on the coordinates $x_a$ adapted to this
surface.

Let $$ g=\left( \begin{array}{cc} V & W \\ W & X
\end{array} \right),\; \rho^2=\det g.
$$
Then, the first group of Einstein equations can be written as
\begin{equation}
d\ast\rho d g g^{-1}=0.\label{EinsteinEq1}
\end{equation}
Here $\ast$ is the Hodge operator with respect to the metric
$ds_{II}^2$. The function $\rho$ has a space-like gradient hence
it can be used as a radial coordinate. Let the time coordinate be
dual to space one, $dt=\ast d\rho$. Coordinate chart constructed
is often called the Weyl canonical coordinates. We employ it
throughout the paper. In these coordinates the 2-metric
$ds_{II}^2$ is given by
\begin{equation}
ds_{II}^2=f(\rho,t)\left(-dt^2+d\rho^2\right),\;\;
f(\rho,t)=\frac{X}{\rho^2}e^{-2\gamma}
\end{equation}
The last formula is a definition of the function $\gamma$. Then
the system (\ref{EinsteinEq1}) takes the form
\begin{equation}
-(\rho g_{,t}g^{-1})_{,t}+(\rho
g_{,\rho}g^{-1})_{,\rho}=0.\label{EinsteinEq1a}
\end{equation}
We impose the following boundary conditions on $g$
\begin{equation}
g=\left(\begin{array}{cc}V&\rho^2\hat W\\ \rho^2\hat W&\rho^2\hat
X\end{array}\right)\label{bc1}
\end{equation}
where $V, \hat W$ and $\hat X$ are smooth functions not equal to
zero and $$V=1+J/\rho+O(1/\rho^2),\;\hat
X=1-J/\rho+O(1/\rho^2),\;\hat W=O(1/\rho^3)$$ at $\rho=\infty$.
Here $J$ is a constant independent on $t$.

An alternative formulation of Einstein equations
(\ref{EinsteinEq1}) exists. For this, let us introduce matrix
potential
\begin{equation}
dH=\rho\ast dgg^{-1},\;\;Y=H_{12}.
\end{equation}
Here $Y$ is the Ernst potential that is determined by the
transition Killing vector field. The functions $Y$ and $V$ are
independent dynamical variables which set a solution of
(\ref{EinsteinEq1}) completely. Potentials $Y$ and $V$ satisfy the
Ernst equation which we will not use in this paper.

Using (\ref{bc1}) one can easily prove that
\begin{equation}
\rho g_{,\rho}g^{-1}=\left(\begin{array}{cc}0&\partial_t Y\\
0&2\end{array}\right),\;\;\rho g_{,t}g^{-1}=0\label{bc-a}
\end{equation}
at $\rho=0$ and
\begin{equation}
\rho g_{,\rho}g^{-1}-
\left(\begin{array}{cc}0&0\\0&2\end{array}\right)=
\left(\begin{array}{cc}O(1/\rho)&O(1/\rho^3)\\
O(1/\rho)&O(1/\rho)\end{array}\right),
\end{equation}
\begin{equation}
\rho g_{,t}g^{-1}= \left(\begin{array}{cc}O(1/\rho)&O(1/\rho^3)\\
O(1/\rho)&O(1/\rho)\end{array}\right),
\end{equation}
at $\rho=\infty$.

The second group of Einstein equations allows one to determine the
coefficient $\gamma$ from the matrix $g$. From (\ref{bc1}) it is
possible to show that $\partial_t\gamma=0$ as $\rho=0$. The
solution is regular at the symmetry axis iff $\gamma=0$ for
$\rho=0$. However, in this case $\gamma\to\gamma_0\ne0$ as
$\rho\to\infty$ and the conical singularity appears at the space
infinity. Its angle of deficit can be treated as the energy of the
system \cite{probing}. It is interesting to note that the deficit
angle of the conical singularity at the symmetry axis of the
stationary solutions with disconnected horizon has the sense of
the force between the black holes (see Ref.\cite{mine} and
references therein). In the present paper, we restrict ourselves
to the study of the system (\ref{EinsteinEq1a}).
\end{section}

\begin{section}{Auxiliary linear problem}

System of equations (\ref{EinsteinEq1a}) is the compatibility
condition for the following pair of matrix linear differential
equations:
\begin{equation}
D_1\Psi=U(\omega,\rho,t)\Psi,\;\;
U(\omega,\rho,t)={\rho^2g_{,\rho}g^{-1}+\omega\rho
g_{,t}g^{-1}\over \rho^2-\omega^2},\label{UV1}
\end{equation}
\begin{equation}
D_2\Psi=V(\omega,\rho,t)\Psi,\;\;
V(\omega,\rho,t)={\rho^2g_{,t}g^{-1}+\omega\rho
g_{,\rho}g^{-1}\over \rho^2-\omega^2},\label{UV2}
\end{equation}
Here $D_1$ and $D_2$ are the commuting differential operators
\begin{equation}
D_1=\partial_\rho-{2\omega\rho\over\omega^2-
\rho^2}\partial_\omega,
\;\;D_2=\partial_t-{2\omega^2\over\omega^2-\rho^2}
\partial_\omega,
\end{equation}
and $\omega$ is a complex parameter that does not depend on $\rho,
t$. We also use the $U-V$-pair representation in which $\omega$ is
a dependent parameter. To be more precise, let $\omega$ be a root
of the equation
\begin{equation}
\omega^2+2(t-k)\omega+\rho^2=0 \label{onomega}
\end{equation}
where $k$ is an independent parameter. Determined through
(\ref{onomega}), $\omega$ satisfies
\begin{equation}
\partial_\rho\omega={2\omega\rho\over \rho^2- \omega^2},\;\;
\partial_t\omega={2\omega^2\over \rho^2-
\omega^2}.
\end{equation}
Passing from $\Psi(\omega)$ to $\Psi^\prime(k)=\Psi(\omega(k))$,
one obtains from (\ref{UV1}, \ref{UV2}) that
\begin{equation}
\partial_\rho\Psi^\prime=U(\omega(k),\rho,t)\Psi^ \prime,\;\;
\partial_t\Psi^\prime=V(\omega(k),\rho,t)\Psi^\prime.\label{UVink}
\end{equation}
We will denote the solution of (\ref{UVink}) by $\Psi(k)$ missing
the prime for brevity. It is worth mentioning that solving
(\ref{UVink}) with the fixed branch of the root one finds the
solution of (\ref{UV1}, \ref{UV2}) only in the analyticity domain
of $\omega(k)$.

We will follow the general scheme for investigating integrable
equations Ref.\cite{T-F} omitting many technical details.

Eq. (\ref{onomega}) is invariant w. r. t. the transformation
$\omega\rightarrow \rho^2/\omega$. Now suppose $\omega_\pm(k)$ are
root's branches such that ${\rm Im} \omega_+>0$ and ${\rm Im}
\omega_-<0$. Then the cuts of $\omega_\pm$ are half-lines
$(-\infty, t-\rho]$ and $[t+\rho,\infty)$. Before proceed we
mention some useful properties of $\omega_\pm$. Note that
\begin{equation}
\omega_+(k)=\frac{\rho^2}{\omega_-(k)},\;\;\;\bar\omega_+(\bar
k)=\omega_-(k).\label{rootsym}
\end{equation}
As $\rho\to\infty$ the functions $\omega_\pm(k,\rho,t)$ satisfy
uniformly in $k, t$ the following estimate
\begin{equation}
\omega_\pm(k,\rho,t)=\pm i\rho+(k-t)+O(1/\rho).\label{rhoinfty}
\end{equation}

Furthermore, introduce the monodromy matrixes
$T_\pm(k,\rho,\rho^\prime)$. By definition, they are solutions to
\begin{equation}
\partial_\rho
T_\pm(\rho,\rho^\prime)=U(\omega_\pm(k),\rho,t)T_\pm(\rho,
\rho^\prime),\;\;T_\pm(\rho,\rho)=I.\label{onT1}
\end{equation}
Besides, zero caveture condition, $$U_{,t}-V_{,\rho}+[U,V]=0,$$
reveals that
\begin{equation}
\partial_t T_\pm(\rho,\rho^\prime)=
V(\omega_\pm(k),\rho)T_\pm(\rho,\rho^\prime)-
T_\pm(\rho,\rho^\prime) V(\omega_\pm(k),\rho^\prime). \label{onT2}
\end{equation}

For symmetric real matrix $g$ the system (\ref{UV1}, \ref{UV2})
remains invariant under the transformations
\begin{equation}
\Psi(\omega,\rho,t)\rightarrow g\tilde\Psi^{- 1}({\rho^2
\over\omega},\rho,t),\;\;\Psi(\omega,\rho,t)\rightarrow
\bar\Psi(\bar\omega,\rho,t)\label{psisym}
\end{equation}
where the tilde denotes transposition. Taking into account
(\ref{rootsym}) one gets
\begin{equation}
T_+(k,\rho,\rho^\prime)=g(\rho)\tilde T_-^{-
1}(k,\rho,\rho^\prime)
g^{-1}(\rho^\prime),\;\;T_+(k,\rho,\rho^\prime)= \bar T_-(\bar
k,\rho,\rho^\prime)\label{Tsym}
\end{equation}
Note that we regard the parameter $k$ as a parameter on the
complex plane, not on the Riemann surface of the root, and
understand equations (\ref{onT1}) as two different ones. Let us
remark also that $T_+(\omega,\rho,\rho^\prime)$ and
$T_-(\omega,\rho,\rho^\prime)$ are solutions of (\ref{UV1}) as
${\rm Im} \omega\geq 0$ and ${\rm Im} \omega\leq 0$ respectively.

Let us determine the Jost functions,
\begin{equation}
\Psi_\pm(k,\rho)=\lim\limits_{\rho'\to\infty}
T_\pm(k,\rho,\rho')e_\pm(k,\rho'),\;e_\pm(k,\rho')=\left(
\begin{array}{cc}1&0\\0&\omega_\pm(k,\rho')\end{array}\right).
\label{jostf}
\end{equation}
They are analytic in the $k$-plane with the cuts
$(-\infty,t-\rho]$ and $[t+\rho,\infty)$. Moreover, from equation
(\ref{onT2}) we derive that they are solutions of compatible
system of equations (\ref{UVink}). Since
$\lim_{\rho\to\infty}e_-(\rho)g^{-1}(\rho)e_+(\rho)=I$ we see also
that
\begin{equation}
\Psi_+(k)=g\tilde\Psi_-^{-1}(k),\;\;\Psi_+(k)=\bar\Psi_-(\bar
k).\label{PsiSym}
\end{equation}
It is worth mentioning that $\det\Psi_\pm(k)=\omega_\pm(k)$.

Recall that functions $\Psi_+(\omega)$ and $\Psi_-(\omega)$ are
{\it analytic} solutions of (\ref{UV1}), (\ref{UV2}) as
$\mbox{Im}\omega>0$ and as $\mbox{Im}\omega<0$ respectively. Hence
for real $\omega$ there is a matrix $T(k)$ such that
\begin{equation}
\Psi_+(\omega)=\Psi_-(\omega)T(k),\;\;\;\;
k=t+\frac{\omega^2+\rho^2}{2\omega}. \label{psiRHP}
\end{equation}
Since $\Psi_\pm(\omega)$ are coupled by transformation
(\ref{psisym}) and $\det\Psi_\pm(\omega)=\omega$, $T(k)$ satisfies
\begin{equation}
T(k)=\tilde T(k),\;\;\;\bar T(\bar k)=T^{-1}(k),\;\;\;\det T(k)=1,
\end{equation}
by the other words it is a symmetric unitary matrix. It is
interesting to note that if initial data $(V-1, \hat X-1,
V_{,t},\hat X_{,t})$ are of compact support then $T-I$ is also of
compact support. In general case we have $T(k)=I+O(1/|k|)$ as
$|k|\to\infty$.

Let
\begin{equation}
\chi_\pm(\omega)=\Psi_\pm(\omega)e^{-1}(\omega),\;\;\;e(\omega)=
\left(\begin{array}{cc}1&0\\0&\omega\end{array}\right).
\end{equation}
It is possible to prove that $\chi_\pm$ are regular at $\omega=0$
and $\chi_\pm(\infty)=I$. The last condition and symmetry
(\ref{psisym}) reveal that
\begin{equation}
g=\chi_\pm(0)\left(\begin{array}{cc}1&0\\0&\rho^2\end{array}\right)
\end{equation}
Since $\chi_-(0)=\chi_+(0)$ from (\ref{psiRHP}) we derive that
$T_{12}(k)=T_{21}(k)=O(1/|k|^2)$ at $k=\infty$.

Boundary condition (\ref{bc-a}) yields
$$V(\omega)|_{\rho=0}=-\frac{1}{\omega}\left(\begin{array}{cc}0&\partial_t
Y\\ 0&2\end{array}\right)$$ Solving (\ref{UV2}) one obtains
\begin{equation}
\Psi_\pm(\omega)|_{\rho=0}=\left(\begin{array}{cc}1&-Y(t)\\0&\omega\end{array}\right)T_\pm(k),
\;\;\;T_-(k)=\bar T_+(\bar k)\label{psiat0}
\end{equation}
Here $T_+$ is analytic in $k$ as $\mbox{Im}k>0$. Inserting
(\ref{psiat0}) in (\ref{psiRHP}) we come to identities:
\begin{equation}
T_-^{-1}(k)T_+(k)=T(k)=\tilde T_+(k)\tilde T_-^{-1}(k)\label{Trep}
\end{equation}
It follows from normalization of $\chi_\pm$ at $\omega=\infty$
that
\begin{equation}
\lim\limits_{\omega\to\infty}e(\omega)T_+(k)e^{-1}(\omega)=I
\label{NorOfTplus}
\end{equation}

As a function of $\omega$, $T_+(k(\omega))$ is analytic in
domains: $\mbox{Im}\omega>0, |\omega|>\rho$ and
$\mbox{Im}\omega<0, |\omega|<\rho$. Then using (\ref{psiRHP}) and
(\ref{Trep}) one can easily check that
\begin{equation}
\Phi_-(\omega)=\left\{\begin{array}{cc}\Psi_+(\omega)T_+^{-1}(k)\;\;\;
\mbox{Im}\omega>0, |\omega|>\rho \\
\Psi_-(\omega)T_-^{-1}(k)\;\;\; \mbox{Im}\omega<0,
|\omega|>\rho\end{array}\right.
\end{equation}
is analytic as $|\omega|>\rho$ while
\begin{equation}
\Phi_+(\omega)=\left\{\begin{array}{cc}\Psi_+(\omega)\tilde
T_-(k)\;\;\; \mbox{Im}\omega>0, |\omega|<\rho \\
\Psi_-(\omega)\tilde T_+(k)\;\;\; \mbox{Im}\omega<0,
|\omega|<\rho\end{array}\right.
\end{equation}
is analytic as $|\omega|<\rho$. On the circle $|\omega|=\rho$,
$\Phi_\pm$ satisfy the conjugation condition,
\begin{equation}
\Phi_+(\omega)=\Phi_-(\omega)G(k),\;\;\;G(k)=T_+(k)\tilde
T_-(k)=T_-(k)\tilde T_+(k).\label{PhiRHP}
\end{equation}
Set $\hat\chi_\pm=\Phi_\pm e^{-1}$. The matrices $\Phi_+$ and
$\Phi_-$ are still coupled by transformation (\ref{psisym}). Using
it and the fact that $\hat\chi_-(\infty)=I$ which follows from
(\ref{NorOfTplus}), we come to
\begin{equation}
g=\hat\chi_+(0)\left(\begin{array}{cc}1&0\\0&\rho^2\end{array}\right)
\end{equation}

Due to the boundary condition (\ref{bc1}) the limit,
$\lim_{\rho\to0}\hat\chi_+(0)$, exists. On the other hand we have
from (\ref{PhiRHP}) that $$
\lim\limits_{\rho\to0}\hat\chi_+(0)=\lim\limits_{\omega\to0}
\left(\begin{array}{cc}1&-Y(t)\\0&\omega\end{array}\right)G(k)e^{-1}(\omega).
$$ The above limit exits if and only if
$Y(t)=G_{12}(t)/G_{22}(t)$. Eventually,
\begin{equation}
\left(\begin{array}{cc}V(t)&\hat W(t)\\0&\hat
X(t)\end{array}\right)=\left(\begin{array}{cc}\frac{1}{G_{22}(t)}&
\frac{2\partial_tG_{12}(t)G_{22}(t)-2G_{12}(t)\partial_tG_{22}(t)}{G_{22}(t)}
\\0&G_{22}(t)\end{array}\right)\label{symaxistoG}
\end{equation}
where unimodularity of $G(k)$ was used as well.

We summarize the results of this section as follow. For any
solution of the initial-value problem posed in Section 2 there
exits analytic in upper $k$-plane matrix, $T_+(k)$, that satisfies
the conditions, $$T_+T_+^\dag=T_+^\dag T_+,\;\;\
\mbox{Im}T_+T_+^\dag=0$$ for real $k$. Here $\dag$ is the
Hermitian conjugate. It uniquely determines symmetric unitary
matrix, $T(k)$, and symmetric real matrix, $G(k)$. Then the
inverse problem is reduced to the Riemann-Hilbert problem (RHP)
(\ref{psiRHP}) or (\ref{PhiRHP}). In addition, there is one to one
correspondence between $G(k)$ and the data on the symmetry axis,
for example $Y(t)$ and $V(t)$ (\ref{symaxistoG}).

\end{section}

\begin{section}{Einstein-Rosen waves}

In this section we apply the results of the previous section to a
space-time with diagonal metric ($W=0$). Set $X=\rho^2e^{-2\psi}$.
Then the system (\ref{EinsteinEq1a}) reduces to one linear
equation, viz
\begin{equation}
-\frac{\partial^2\psi}{{\partial
t}^2}+\frac{\partial^2\psi}{{\partial\rho}^2}+
\frac{1}{\rho}\frac{\partial\psi}{\partial\rho}=0
\label{EinsteinEq1aPsi}
\end{equation}
The above equation is the compatibility condition of the following
pair of scaler linear equations
\begin{equation}
\partial_\rho F=\frac{2\rho^2\partial_\rho\psi+
2\rho\omega_\pm\partial_t\psi}{\rho^2-\omega_\pm^2}F,\;\;
\partial_t F=\frac{2\rho^2\partial_t\psi+
2\rho\omega_\pm\partial_\rho\psi}{\rho^2-\omega_\pm^2}F.
\end{equation}
The solution of the system (\ref{UVink}) is given by
$$\Psi_\pm(k,\rho,t)=\left(\begin{array}{cc}e^{\theta_\pm(k,\rho,t)}&0
\\0&\omega_\pm
e^{-\theta_\pm(k,\rho,t)}\end{array}\right)$$ with
$$\theta_\pm(k,\rho,t)=-\int\limits_\rho^\infty
d\rho'\left(\frac{2\rho'^2\partial_\rho\psi}{\rho'^2-\omega_\pm^2}+
\frac{2\rho'\omega_\pm\partial_t\psi}{\rho'^2-\omega_\pm^2}\right).$$

The function $k(\omega)=t+\frac{\omega^2+\rho^2}{2\omega}$ maps
the upper half-plane(lower half-plane) into $k$-plane with the
cuts $(-\infty,t-\rho]$ and $[t+\rho,\infty)$. The function
$\omega_+(k,\rho')(\omega_-(k,\rho'))$ maps the $k$-plane with the
cuts $(-\infty,t-\rho']$ and $[t+\rho',\infty)$ into the upper
half-plane(lower half-plane). Analyzing these mappings in the case
when $\rho\leq\rho'$ we conclude that $$
\omega_+(k(\omega+i0),\rho')-\omega_-(k(\omega-i0),\rho')=
2i\sqrt{\rho'^2-(k-t)^2} $$ as $|k-t|\leq\rho'$ and zero
otherwise. Note that $|k-t|\geq\rho$ for any real $\omega$ and
$\omega_\pm(k,\rho')=k-t\pm i\sqrt{\rho'^2-(k-t)^2}$ as
$\rho\leq|k-t|\leq\rho'$ and real $k$. Therefore
\begin{equation}
\theta_+(k(\omega+i0),\rho,t)-\theta_-(k(\omega-i0),\rho,t)=2i\Delta(k)
\label{sRHP}
\end{equation}
where $$ \Delta(k)=-\int\limits_{|k-t|}^\infty d\rho'
\frac{(k-t)\partial_\rho\psi+\rho'\partial_t\psi}{\sqrt{\rho'^2-(k-t)^2}}.
$$

The piece-wise analytic function $\theta(\omega)$,
$\theta(\omega)=\theta_-(\omega)$ as $\mbox{Im}\omega<0$ and
$\theta(\omega)=\theta_+(\omega)$ as $\mbox{Im}\omega>0$, is a
unique solution of the RHP (\ref{sRHP}) with $\theta(\infty)=0$.
This solution can be written as
\begin{equation}
\theta(\omega)=\frac{1}{\pi}\int\limits_{-\infty}^\infty
d\lambda\frac{\Delta(k(\lambda))}{\lambda-\omega}.
\end{equation}
Then we can represent the solution of (\ref{EinsteinEq1aPsi}) as
\begin{equation}
\psi(t,\rho)=\frac{1}{2}\theta(0)=\frac{1}{2\pi}\int\limits_{-\infty}^\infty
d\lambda\frac{\Delta(k(\lambda))}{\lambda}.
\end{equation}
Changing the variable in the above integral one has
\begin{equation}
\psi(t,\rho)=\frac{1}{\pi}\int\limits_{-\infty}^{t-\rho}dk
\frac{\Delta(k)}{\sqrt{(k-t)^2-\rho^2}}+
\frac{1}{\pi}\int\limits_{t+\rho}^\infty dk
\frac{\Delta(k)}{\sqrt{(k-t)^2-\rho^2}}
\end{equation}
\end{section}

\section{Conclusions}

In this paper we outline the ISM scheme for the cylindrical
symmetric waves model. However, to complete this scheme one needs
to analyze the smoothness property of the matrices $T(k)$ and
$G(k)$. We assume that for smooth initial data they are also
smooth. The relation of the matrix $G$ with the data on the
symmetry axis allows to determine the behavior of the solution at
the time infinities through the study of the asymptotic properties
of $G$. The matrix $T$ defines the behavior of the solution at the
null infinities and we assume that it is uniquely restored from
the data on the null infinities. An important open question is the
derivation of the formula for the angle of deficit at the space
infinity in terms of the matrix $T$ or $G$. We plan to consider
the problems mentioned in separate papers without promising to do
it soon.

\section*{Acknowledgements}
This work was partly supported by RFBR grant No 98-01-01063.

\end{document}